\documentclass[twocolumn]{jpsj3} 

\title{Quantum Phase Transition in Antiferromagnetic Heisenberg Chains Coupled to Phonons}

\author{Satoru \textsc{Akiyama}\thanks{E-mail address:akiyama@wakayama-nct.ac.jp} 
 and Chitoshi \textsc{Yasuda}$^{1}$\thanks{E-mail address:cyasuda@phys.u-ryukyu.ac.jp}}

\inst{Department of General Education, Wakayama National College of Technology, Noshima 77, Nada, Gobo, Wakayama 644-0023  \\
$^{1}$Department of Physics and Earth Sciences, Faculty of Science, University of the Ryukyus, 1 Senbaru, Nishihara, Okinawa 903-0213}

\abst{We develop an analytical approach based on a unitary transformation to investigate $S=1/2$ antiferromagnetic Heisenberg
chains coupled to phonons, and find a new quantum phase transition at zero temperature.
Although the usual phase transition occurs depending on the strengths of the spin-spin and spin-phonon interactions, 
the phase transition in the present work is induced by a change in the geometrical structure of the spin-phonon interaction.
The magnetic properties of the phase transition can be understood in relation to a phase transition 
that has already been found in the frustrated $J_1$-$J_2$ model.
}

\recdate{\today}

\kword{spin-phonon interaction, frustration, effective Hamiltonian, level spectroscopy, exact diagonalization, 
quantum phase transition, spin-Peierls compound}

\begin{document}
\maketitle

\section{Introduction} 

Quantum spin systems with spin-phonon interactions have actively been studied 
as model systems for the research on
spin-Peierls compounds, particularly the first inorganic spin-Peierls compound CuGeO$_3$.
In the spin-Peierls compounds, a dimerized state with a spin gap due to a topological structure is realized 
at low temperatures~\cite{hase}. 
It results from strong quantum fluctuations for quasi-one-dimensional systems~\cite{pytte,cross,nakano}.
In the one-dimensional systems, it has been theoretically elucidated that a spin-liquid state or a dimerized state is
realized as the ground state depending on the strengths of the spin-spin and spin-phonon interactions, 
and the phase transition between the states belongs to the same universality class as that of the 
Berezinskii-Kosterlitz-Thouless-type~\cite{BKT} transition in the $J_1$-$J_2$ model~\cite{bursill}.

The quantum spin systems with the spin-phonon interactions have theoretically been studied using 
effective spin Hamiltonians~\cite{kuboki,uhrig,weise}.
In the previous works, the effective spin Hamiltonians are described by the $J_1$-$J_2$ model or the frustrated models
with more long-range interactions.
The Hamiltonian of the $J_1$-$J_2$ model is
\begin{equation}
 {\cal H} = J_1 \sum_i \mib{S}_i \cdot \mib{S}_{i+1} + J_2 \sum_i \mib{S}_i \cdot \mib{S}_{i+2} \ ,
 \label{j1j2}
\end{equation}
where $\mib{S}_i$ is the spin operator on site $i$. The values of $J_1$ and $J_2$ are positive
and, therefore, the first and second terms on the right-hand side of eq.~(\ref{j1j2}) are competitive
interactions. In such a case, stable spin states are not easily found owing to the strong frustration.
In a special case of $J_2=J_1/2$, the ground state is exactly the product of spin-singlet states
(Majumdar-Ghosh point)~\cite{majumdar}. The phase transition 
between the spin-liquid phase and the dimer phase in the $J_1$-$J_2$ model 
has been found by the level spectroscopy technique that considers
the numerical data obtained by the exact diagonalization method~\cite{nomura}.
In the quantum spin systems with the spin-phonon interactions, besides these results, 
the following results have been obtained: both effects of phonons and frustration cooperatively work 
to form the gapped phase~\cite{weise,augier}, and the separation of time scales of quantum fluctuations of spin 
and lattice degrees of freedom (the quantum narrowing effect) has been found by quantum Monte
Calro simulation~\cite{onishi}.

\begin{figure}[t]
  \centerline{\resizebox{0.45\textwidth}{!}{\includegraphics{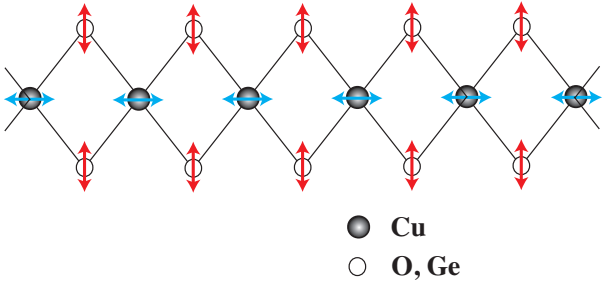}}}
  \caption{Illustration of lattice distortions for CuGeO$_3$. The spin-phonon interactions on Cu spins are induced by two types of displacement for Cu itself and O or Ge atoms.}  
\label{lattice}
\end{figure}

In the present work, we have developed an analytical approach by which the phonon degrees of freedom 
are trace-outed and 
an effective Hamiltonian with frustration is yielded.
Applying this method to a one-dimensional system with a more general spin-phonon interaction than usual
cases of the previous works, we have found the ground-state phase transition between the spin-liquid 
phase and the dimer phase. 
In the usual cases, the spin-phonon interaction with difference coupling~\cite{bursill,kuboki,uhrig,weise,hager},
\begin{equation}
 \label{sp_diff}
   {\cal H}_{\rm sp}^{\rm diff} = g \sum_i (b_i^{\dagger}+b_i)(\mib{S}_i \cdot \mib{S}_{i+1} - \mib{S}_i \cdot \mib{S}_{i-1}) \ ,
\end{equation}
and/or that with local coupling~\cite{weise,augier,onishi,wellein,weise2},
\begin{equation}
 \label{sp_loc}
   {\cal H}_{\rm sp}^{\rm loc} = g \sum_i (b_i^{\dagger}+b_i)\mib{S}_i \cdot \mib{S}_{i+1}  \ ,
\end{equation}
have been investigated, where $b_i^{\dagger}$ and $b_i$ are the phonon creation and annihilation operators at site $i$, respectively,
and $g$ is the coupling constant.
The difference coupling was well studied before the discovery of CuGeO$_3$ and 
the local coupling was well studied in the research on CuGeO$_3$.
The magnetism of CuGeO$_3$ is due to Cu atoms, and  
the crystal structure shows that the spin-phonon interaction is described by~\cite{werner}
\begin{equation}
  \label{cugeo3}
   {\cal H}_{\rm sp} = \sum_i [ g^{\rm Cu} (b_i^{\dagger}+b_i-b_{i+1}^{\dagger}-b_{i+1}) + g^{\rm O, Ge} (b_i^{\dagger}+b_i) ]
   \mib{S}_i \cdot \mib{S}_{i+1} \ ,
\end{equation}
where $\mib{S}_i$ denotes the spin of a Cu atom.
As shown in Fig.~\ref{lattice}, the spin-phonon interactions on Cu spins are induced
by two types of displacement for Cu itself and O or Ge atoms.
The term $g^{\rm Cu}$ of eq.~(\ref{cugeo3}) indicates the spin-phonon interaction caused by the displacement of Cu atoms, and 
the term $g^{\rm O, Ge}$ is 
caused by the displacement of O or Ge atoms.
Generally, although the displacements are written in terms of the different boson operators,
we consider a simple mode: the Cu, O, and Ge atoms link and move, and the displacements are written in terms of one boson operator.
In the research on CuGeO$_3$, the interaction
without the term $g^{\rm Cu}$, i.e., the local coupling, is used as the relevant model because $g^{\rm O, Ge} \gg g^{\rm Cu}$~\cite{werner,affleck}.

In the present work, we have studied the $S=1/2$ spin-phonon interaction described by
\begin{equation}
 \label{sp}
   {\cal H}_{\rm sp} = g \sum_i (b_i^{\dagger}+b_i)(\mib{S}_i \cdot \mib{S}_{i+1} - \beta \mib{S}_i \cdot \mib{S}_{i-1}) \ .
\end{equation}
This interaction corresponds to interaction~(\ref{cugeo3}) through the relations
\begin{equation}
  \label{beta}
    \beta = \frac{g^{\rm Cu}}{g^{\rm Cu}+g^{\rm O, Ge}} \ ,
\end{equation}
and $g=g^{\rm Cu}+g^{\rm O, Ge}$.
When $g^{\rm O, Ge} \ll g^{\rm Cu}$ and $g^{\rm O, Ge} \gg g^{\rm Cu}$, $\beta=1$ and 0, respectively. 
Therefore, this interaction
includes both limits of the difference coupling~(\ref{sp_diff}) and the local coupling~(\ref{sp_loc}).
In the present work, we have focused on eq.~(\ref{sp}) as the spin-phonon interaction and found
the ground-state phase transition between the spin-liquid phase and the dimer phase by changing the 
value of $\beta$.
Although the phase transition is similar to that well-known in the system with usual 
interactions~(\ref{sp_diff}) and (\ref{sp_loc}), it is a new phase transition in the sence 
that the driving force is $\beta$ not $g$.

The present paper is organized as follows. In \S 2, the analytical approach based on a unitary transformation is introduced and 
the effective Hamiltonians are obtained using the Ising and Heisenberg models coupled to phonons. 
In \S 3, numerical results for the $S=1/2$ antiferromagnetic Heisenberg chains coupled to phonons are presented.
Section 4 is devoted to summary and discussion.

\section{Effective Hamiltonian} 

We consider the Hamiltonian described by 
\begin{eqnarray}
h &=& h_{\rm p} + t h_{\rm s} + g h_{\rm sp} \ ,
\label{eq:h}\\
h_{\rm p} &=& \sum_{i = 1}^{N} b_{i}^{\dagger} b_{i} \ , 
\label{eq:hp}\\
h_{\rm sp} &=& \sum_{i = 1}^{N} (b_{i}^{\dagger}+b_{i})~v_{i} \ ,
\label{eq:hsp}
\end{eqnarray}
where $N$ is the number of spin sites, $g$ is the coupling constant of the spin-phonon interaction, and
$t$ is the positive exchange integral of the antiferromagnetic spin-spin interaction. 
The phonon frequency $\omega$ is considered a unity, i.e., we assume an 
antiadiabatic case.  
The terms $h_{\rm s}$ and $v_i$ have only to satisfy the conditions $[h_{\rm p}, h_{\rm s}]=[h_{\rm p}, v_i]=0$.
We consider the following unitary transformation~\cite{weise} for Hamiltonian~(\ref{eq:h}):
\begin{eqnarray}
\bar{h}
&=& e^{g {\cal S}} h e^{- g {\cal S}}
= \sum_{k = 0}^{\infty} \frac{g^{k}}{k!} [{\cal S}~,~h]_{k} \nonumber \\
&=& \sum_{k = 0}^{\infty} \frac{g^{k}}{k!} \left(
	[{\cal S}~,~h_{\rm p} + t h_{\rm s}]_{k}
	+ k [{\cal S}~,~h_{\rm sp}]_{k-1}
\right) \ ,
\label{eq:hbar}
\end{eqnarray}
where ${\cal S}$ is an anti-Hermite operator of the order of $g^0$ and $[A,~B]_{k} = \underbrace{[A,~[A,~\dots~[A,}_{k}~B]~]]$.
Wei$\beta$e {\it et al}.~\cite{weise} determined the form of ${\cal S}$ so that
`the amplitude of the first-order term in $g$ of $\bar{h}$ operated to eigen states of the zeroth-order term in $g$ is minimal.' 
In the present work, the specific form of ${\cal S}$ is determined so that `the first-order term in $g$ 
of $\bar{h}$ becomes zero':
\begin{equation}
[{\cal S}~,~h_{\rm p} + t h_{\rm s}] + h_{\rm sp} = 0 \ .
\label{eq:g1term}
\end{equation}
As a result, we could find the formal solution of eq.~(\ref{eq:g1term}),
\begin{equation}
{\cal S} = [h_{\rm p} - t h_{\rm s}~,~{\cal K}] \ , 
\label{eq:s2sol}
\end{equation}
with
\begin{equation}
{\cal K} = \sum_{n=0}^{\infty} t^{2n} [h_{\rm s}~,~h_{\rm sp}]_{2n} \ .
\end{equation}
Then, we obtain the transformed Hamiltonian
\begin{equation}
\bar{h} = \sum_{k = 0}^{\infty} \frac{g^{k}}{k!}
	(1 - k) [{\cal S}~,~h_{\rm p} + t h_{\rm s}]_{k}  \ .
\label{eq:hbar1}
\end{equation}

\subsection{Ising model}

First, for a simple example, we consider the one-dimensional Ising model:
\begin{equation}
h_{\rm s} = \sum_{i} \left(
	S^{z}_{i} S^{z}_{i+1} + \alpha S^{z}_{i} S^{z}_{i+2}
\right) \ , 
\end{equation}
and
\begin{equation}
v_{i} = S^{z}_{i} S^{z}_{i+1} - \beta S^{z}_{i} S^{z}_{i-1} \ .
\label{eq:v_isng}
\end{equation}
In this case, the transformed Hamiltonian is exactly obtained as
\begin{equation}
\bar{h} = h_{\rm p} + t \sum_i (S_i^z S_{i+1}^z + \alpha_{\rm eff} S_i^z S_{i+2}^z) - \frac{N}{16}g^2(1+\beta^2) \  ,
\label{eq:hbar_Ising}
\end{equation}
where $\alpha_{\rm eff}=\alpha+\frac{g^2 \beta}{2t}$.
The second term on the right-hand side of eq.~(\ref{eq:hbar_Ising}) is the $J_1$-$J_2$ Ising model. Its ground state is
the N\'{e}el state for $0 \le a_{\rm eff} < 1/2$
and the uudd state for $\alpha_{\rm eff} > 1/2$~\cite{morita}.
Thus, eq.~(\ref{eq:hbar_Ising}) shows that the spin-phonon interaction acts cooperatively to form the uudd state
for $\beta > 0$. When the ground state of original Hamiltonian~(\ref{eq:h}) is expressed by $|\Phi_0 \rangle$, 
we can obtain the expectation value of the lattice distortion 
$\langle \Phi_0 | b_i^{\dagger} + b_i | \Phi_0 \rangle = 2g(1-\beta)S^2$ for the N\'{e}el state. 
This result shows that the system is translated uniformly for $\beta \ne 1$. For the uudd state, on the other hand, 
$\langle \Phi_0 | b_i^{\dagger} + b_i | \Phi_0 \rangle = (-1)^i 2g(1+\beta)S^2$, and the system is alternately distorted.

\subsection{Heisenberg model}

Next, we consider the one-dimensional $S=1/2$ Heisenberg model:
\begin{equation}
v_{i} = \mib{S}_{i} \mib{S}_{i+1} - \beta \mib{S}_{i} \mib{S}_{i-1} \ .
\label{eq:v_heisenberg}
\end{equation}
For simplicity, we set $t=0$. 
The transformed Hamiltonian up to the second order in $g$ conserves the total phonon number. 
In this case, a vacuum state of phonon $|0\rangle_{\rm p}$ is an exact eigen state.
Thus, the effective spin Hamiltonian $h_{\rm eff} = {}_{\rm p}\langle{0}|~\bar{h}~|0\rangle_{\rm p}$ becomes
\begin{eqnarray}
  h_{\rm eff} &=& \frac{g^2}{2}(1+\beta^2) \sum_i (\mib{S}_i \cdot \mib{S}_{i+1} + \frac{\beta}{1+\beta^2} \mib{S}_i \cdot \mib{S}_{i+2})
  \nonumber \\ 
  &-& \frac{3}{16} N g^2 (1+\beta^2) \ .
   \label{og2-ham}
\end{eqnarray}
We call it the $O(g^2)$ Hamiltonian.
In the first term on the right-hand side of eq.~(\ref{og2-ham}), the ratio of the strength of the nearest-neighbor term
to that of the next-nearest-neighbor term, which corresponds to $J_2/J_1$, does not depend on $g$. 
This is the $J_1$-$J_2$ model with an effective frustration:
\begin{equation}
     \alpha_{\rm eff} = \frac{\beta}{1+\beta^2} \ .
 \label{a_eff-beta}    
\end{equation}
For the local coupling with $\beta=0$, $\alpha_{\rm eff}$ is zero 
and the $O(g^2)$ Hamiltonian becomes the Heisenberg model with only the nearest-neighbor interaction. 
For the difference coupling with $\beta=1$, $\alpha_{\rm eff}$ is 1/2 and the $O(g^2)$ 
Hamiltonian describes the Majumdar-Ghosh point. 

The effective Hamiltonian up to the fourth order in $g$ (we call it the $O(g^4)$ Hamiltonian) is
\begin{eqnarray}
h_{\rm eff} &=& J_{0} + J_{1} \sum_{i} \mib{S}_{i} \cdot \mib{S}_{i+1}
+ J_{2} \sum_{i} \mib{S}_{i} \cdot \mib{S}_{i+2} \nonumber \\
&+& J_{3} \sum_{i} \mib{S}_{i} \cdot \mib{S}_{i+3}  \nonumber \\
&+& \Gamma_1  F_{123} + \Gamma_2 F_{213} + \Gamma_3 F_{312}  \nonumber \\
&+& \Gamma_{4} ( F_{214} - F_{412} + F_{324} - F_{423})  \ ,
\label{og4-ham}
\end{eqnarray}
where
\begin{equation}
   F_{\delta_1 \delta_2 \delta_3} \equiv  \sum_i (\mib{S}_{i} \cdot \mib{S}_{i+\delta_1}) (\mib{S}_{i+\delta_2} \cdot \mib{S}_{i+\delta_3}) \ ,
\end{equation}
\begin{eqnarray}
J_{0} &=& -\frac{3}{16} \biggl[ 
	g_1^{2}  
	- \frac{1}{4} g_1^{4} (1 - \alpha_{\rm eff}) (1 + \alpha_{\rm eff}) \biggr] N \ , 
\label{eq:j0}\\
J_{1} &=& \frac{1}{2} \biggl[  
	g_1^{2}   
	- \frac{3}{4} g_1^{4} (1 - \alpha_{\rm eff}) (1 + \alpha_{\rm eff}) \biggr] \ , 
\label{eq:j1}\\
J_{2} &=& \frac{1}{2} \left\{ 
	g_1^{2} \alpha_{\rm eff} - \frac{1}{2} g_1^{4} \biggl[ 
					\alpha_{\rm eff} \right. \nonumber \\ && \hspace{0.3cm} 
					\left. - \frac{3}{4} (1 - \alpha_{\rm eff}) (1 + \alpha_{\rm eff})
	\biggr] 
\right\} \ ,
\label{eq:j2}\\
J_{3} &=& \frac{1}{8} g_1^{4} \alpha_{\rm eff} (2 + \alpha_{\rm eff}) \ ,
\label{eq:j3}\\
\Gamma_{1} &=& -\frac{3}{4} g_1^{4} \alpha_{\rm eff}^{2} \ ,
\label{eq:gammaa}\\
\Gamma_{2} &=& -\frac{1}{4} g_1^{4} \alpha_{\rm eff} (2-3\alpha_{\rm eff}) \ ,
\label{eq:gammab}\\
\Gamma_{3} &=& \frac{1}{2} g_1^{4} \alpha_{\rm eff} (1-\alpha_{\rm eff}) \ ,
\label{eq:gammac}\\
\Gamma_{4} &=& -\frac{1}{4} g_1^{4} \alpha_{\rm eff}^{2} \ ,
\label{eq:gamma1}
\end{eqnarray}
\begin{figure}[t]
  \centerline{\resizebox{0.45\textwidth}{!}{\includegraphics{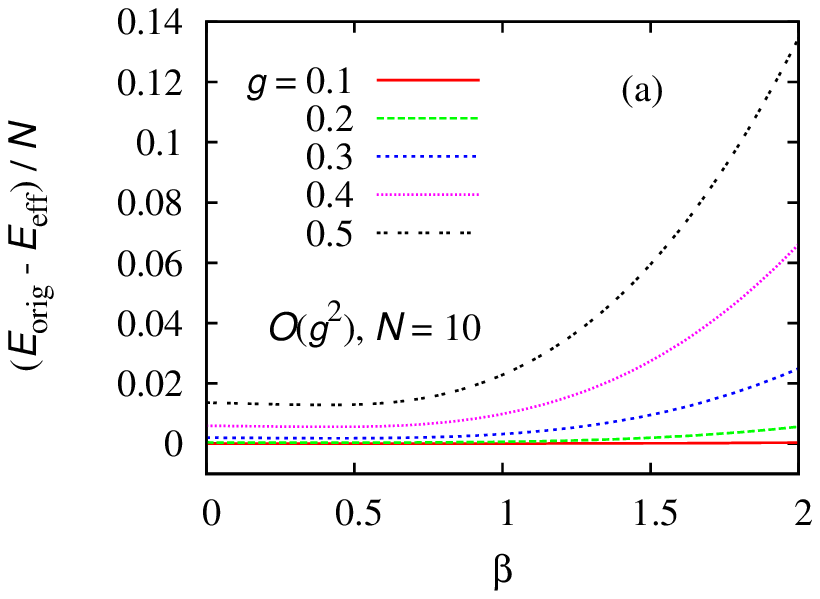}}}
  \centerline{\resizebox{0.45\textwidth}{!}{\includegraphics{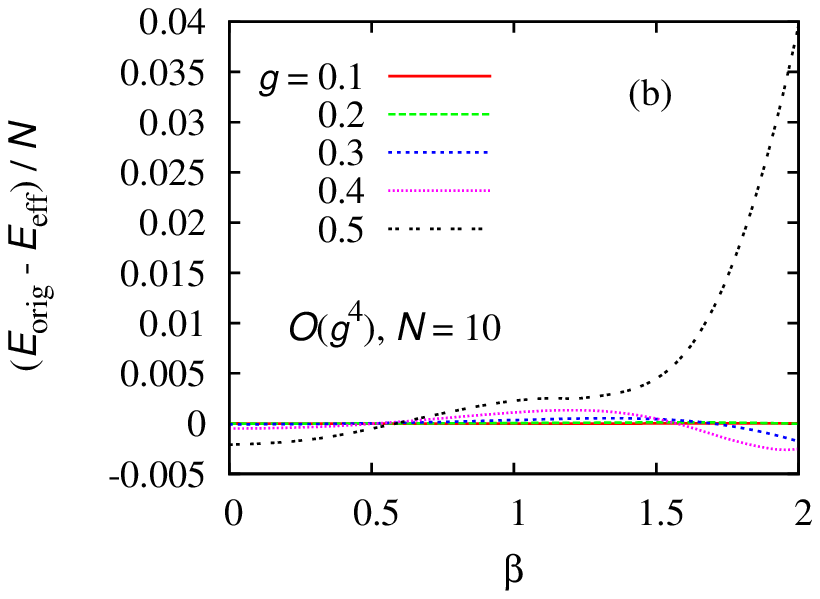}}}  
  \caption{Comparisons between the ground-state energies of the original Hamiltonian and the effective Hamiltonians up to (a) second and
  (b) fourth orders in  $g$ for $N=10$.}  
\label{check-GSE}
\end{figure}
and $g_1 \equiv g \sqrt{1+\beta^2}$. The effective Hamiltonian could be described by $g_1$ and $\alpha_{\rm eff}$ only.
Although the phonon vacuum state $|0\rangle_{\rm p}$ is an exact eigen state of the $O(g^2)$ Hamiltonian,  
this $O(g^4)$ Hamiltonian is an expectation value on $|0\rangle_{\rm p}$. 
The third-nearest spin interaction and the multiple-spin exchange interactions arise from the fourth-order term in $g$.
For $\beta=0$, however, the long-range interactions vanish and the $O(g^4)$ Hamiltonian becomes
the $J_1$-$J_2$ model with the effective frustration:
\begin{equation}
     \alpha'_{\rm eff} = \frac{3g^2}{8-6g^2} \ .
  \label{a_eff-g}
\end{equation} 
In the $J_1$-$J_2$ model, the phase transition between the spin-liquid phase 
and the dimer phase occurs at $\alpha_{\rm c} = 0.2411$~\cite{okamoto}. 
Thus, for the $O(g^4)$ Hamiltonian with $\beta=0$, the phase transition occurs at 
$g_{\rm c} \equiv \sqrt{\frac{8\alpha_{\rm c}}{3+6\alpha_{\rm c}}} \simeq 0.66$.

\begin{figure}[t]
  \centerline{\resizebox{0.45\textwidth}{!}{\includegraphics{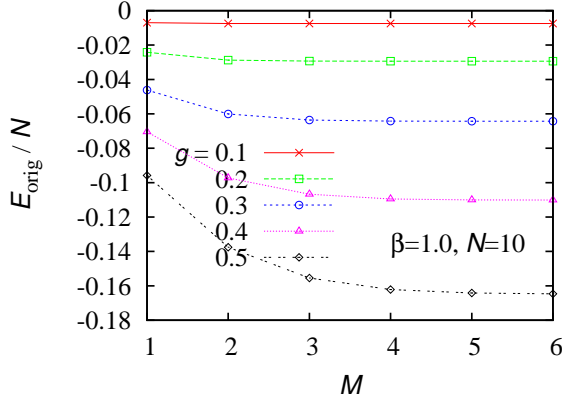}}}   
  \caption{Dependence of the ground-state energy per site for the original Hamiltonian on the cutoff parameter $M$ for $\beta=1$ and $N=10$. The lines serve as visual guides.}  
\label{cutoff}
\end{figure}

\section{Numerical Results}

To determine whether this effective-Hamiltonian method is useful, we show the $\beta$ dependences
of the differences between the ground-state energy per site $E_{\rm orig}/N$ for original Hamiltonian~(\ref{eq:h}) and 
the $E_{\rm  eff}/N$ for effective Hamiltonians~(\ref{og2-ham}) and (\ref{og4-ham}) in  Fig.~\ref{check-GSE}.
The ground-state energies are calculated by the exact-diagonalization method with a periodic boundary condition.
In the original model,
the total Hilbert space is written as the tensorial product space of spins and phonons. The dimension of the subspace associated 
to the phonons is infinite even for a finite chain system. Thus, we apply a truncation procedure retaining only basis states with
the highest number of $M$ phonons, where $M$ is called the cutoff parameter. 
In the present work, we investigated the convergence of the ground-state energy for the truncation procedure.
The dependence of $E_{\rm orig}/N$ on the cutoff parameter is shown in Fig.~\ref{cutoff}. 
The values of $E_{\rm orig}$ in Fig.~\ref{check-GSE} are obtained at $M=6$. 
In Fig.~\ref{check-GSE}(a), we show
results for the $O(g^2)$ effective Hamiltonian. For $g=0.1$, the energies $E_{\rm orig}$ and $E_{\rm eff}$
well agree in the entire range of $0 \le \beta \le 2.0$. 
When $g$ increases, the difference in energy increases for large $\beta$ values. For $g=0.5$, 
the energies disagree in the entire range of $\beta$. In Fig.~\ref{check-GSE}(b), we show results for the $O(g^4)$ effective 
Hamiltonian.  In comparison with the case of $O(g^2)$, the energies well agree in a wider range of $\beta$.
\begin{figure}[t]
  \centerline{\resizebox{0.45\textwidth}{!}{\includegraphics{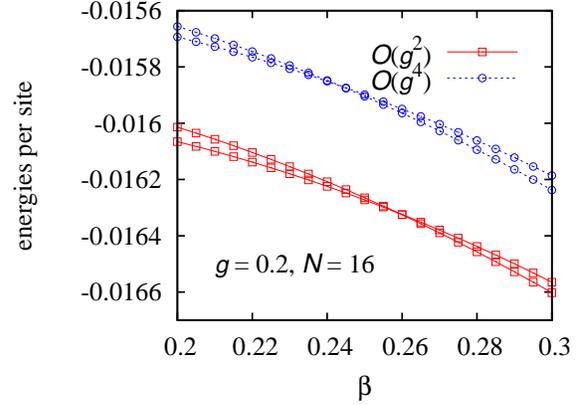}}}
  \caption{$\beta$ dependences of the lowest energy 
  and the first-excited-state energy per site in the $\Sigma_i S_i^z=0$ and $k=\pi$ subspace for $g=0.2$ 
  and $N=16$. The square and circle represent the energies for the effective Hamiltonian up to the second and fourth orders in $g$, respectively.
 The lines serve as visual guides.}  
\label{level cross}
\end{figure}
\begin{figure}[t]
  \centerline{\resizebox{0.45\textwidth}{!}{\includegraphics{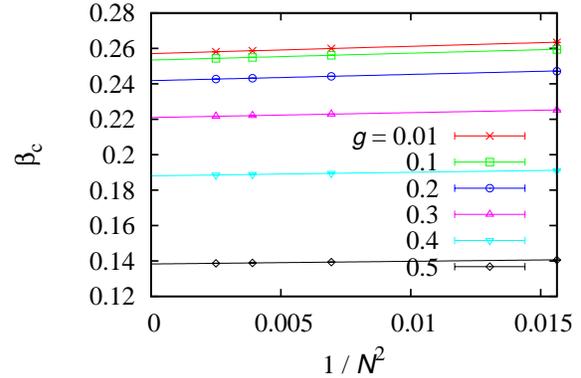}}}  
  \caption{Size dependence of $\beta_{\rm c}(N)$ for various $g$'s for the $O(g^4)$ Hamiltonian. The solid lines 
  are obtained by fitting the data of $N=12$, 16, and 20.
  The values at $1/N^2 = 0$ are the phase-transition points in the thermodynamic limit.}  
\label{fitting}
\end{figure}
\begin{figure}[t]
  \centerline{\resizebox{0.45\textwidth}{!}{\includegraphics{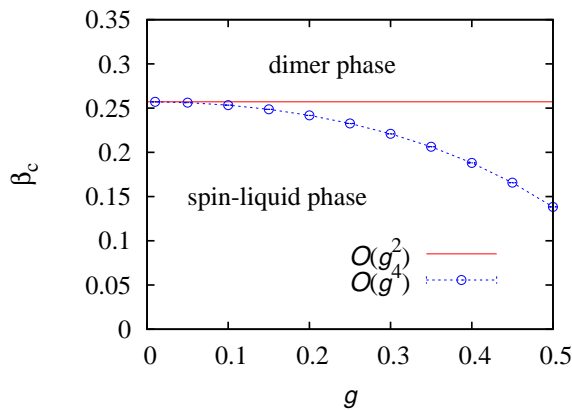}}}  
  \caption{$g$ dependence of the phase-transition point $\beta_{\rm c}$. The solid line corresponds to the transition point between the spin-liquid phase and the dimer phase for the $O(g^2)$ Hamiltonian. 
  The circles represent the results for the $O(g^4)$ Hamiltonian.
  The dashed line serves as a visual guide.}  
\label{beta_c}
\end{figure}

Next, we carry out level-spectroscopy analysis by using the numerical data obtained by the exact diagonalization
of effective Hamiltonians (\ref{og2-ham}) and (\ref{og4-ham}).
Using this method, Nomura and Okamoto~\cite{nomura} have analyzed low-excited states 
of the $S=1/2$ antiferromagnetic XXZ spin chain with 
the next-nearest neighbor interactions and obtained the Berezinskii-Kosterlitz-Thouless-type
phase-transition points from the energy-level crosses. 
As mentioned above, in the $J_1$-$J_2$ Heisenberg model with Hamiltonian (\ref{j1j2}),
the phase-transition point is $J_2/J_1=0.2411$~\cite{okamoto}.
Since our effective $O(g^2)$ Hamiltonian~(\ref{og2-ham}) is the same as the $J_1$-$J_2$ model Hamiltonian,
the phase transition in the present work could also be evaluated by level-spectroscopy analysis.
Although this is not trivial in the $O(g^4)$ Hamiltonian with the multiple-spin-exchange interactions,
we could find an energy-level cross similar to that of the $O(g^2)$ Hamiltonian.  
Both level crosses are obtained using the lowest energy and the first-excited-state energy in the 
$\sum_i S_i^z = 0$ and  $k=\pi$ subspace, which correspond to the dimer and N\'{e}el excitations, respectively~\cite{nomura}.
In Fig.~\ref{level cross}, we show the $\beta$ dependences of the lowest energy and the first-excited-state energy in the subspace.

The phase-transition points in the thermodynamic limit are estimated by the least-squares fitting to the fitting function~\cite{okamoto}
\begin{equation}
 \beta_c(N) = \beta_c + \frac{A}{N^2} \ ,
\end{equation}
where $A$ is a constant number. 
In Fig.~\ref{fitting}, we show the system size dependence of $\beta_c$ for various $g$'s for the $O(g^4)$ Hamiltonian.
The data of $N=12, 16$, and 20 are fitted much better by the fitting function. 

In Fig.~\ref{beta_c}, we show the $g$ dependence of the phase-transition point $\beta_{\rm c}$ between the spin-liquid phase
and the dimer phase obtained by level-spectroscopy analysis.
For the $O(g^2)$ Hamiltonian, naturally, $\beta_{\rm c}$ is constant. The value of 0.2571(1) well agrees with
the solution $\beta \simeq 0.2570$ of the equation $\alpha_{\rm eff}=0.2411=\beta/(1+\beta^2)$ obtained in the $J_1$-$J_2$
model. For large $g$ values, we should consider the higher-order terms in $g$. For the $O(g^4)$ Hamiltonian,
$\beta_{\rm c}$ decreases as $g$ increases. This result shows that the spin-phonon interaction 
acts cooperatively to form the dimerized state.

Thus far, we set $t=0$ for simplicity. If the contribution of the first-order term in $t$ is taken into account 
under the condition $t \simeq g \ll 1$,  
the Hamiltonian $t h_{\rm s} = t \sum_i (\mib{S}_i \cdot \mib{S}_{i+1} + \alpha \mib{S}_i \cdot \mib{S}_{i+2})$ is added to the effective Hamiltonian. Namely, $J_1+t$ and $J_2+t\alpha$ replace $J_1$ and $J_2$, respectively.
Then, the effective frustration becomes
\begin{equation}
  \frac{J_2+t\alpha}{J_1+t} = \frac{J_2}{J_1} + (\alpha - \frac{J_2}{J_1})\frac{t}{J_1+t} \ .
  \label{eff}
\end{equation}
The sign of the second term on the right-hand side of eq.~(\ref{eff}) is positive for $\alpha > J_2/J_1$ and 
negative for $\alpha < J_2/J_1$. Thus, the spin-spin interaction acts cooperatively to form the dimer phase for $\alpha > J_2/J_1$ and
the spin-liqud phase for $\alpha < J_2/J_1$. 

We have also obtained results for the case in which we neglect the four-spin interactions in eq.~(\ref{og4-ham}). 
Since the results well agree with those for the non-neglecting case, the effect of the four-spin interactions is confirmed to be small.

\section{Summary and Discussion}

In the present work, we developed an analytical approach based on a unitary transformation. Applying our method to 
the $S=1/2$ antiferromagnetic Heisenberg chains coupled to phonons, we found a new quantum phase transition at zero 
temperature. 
Although the usual phase transition observed in the difference and local coupling cases occurs 
depending on the strengths of the spin-spin and spin-phonon interactions, the phase transition in the present work is induced by a change in the geometrical structure of the spin-phonon interactions. 
We expect that it  might be experimentally observed in new spin-Peierls compounds or as an effect of pressure. 

In the present work, we consider a nonadiabatic model and set $t=0$ for simplicity.
Neutron scattering experiments have shown that neighboring spins 
within CuGeO$_3$ interact via high-$\omega$ phonons and the interaction is described by the nonadiabatic model 
with $\omega \sim J$, i.e., $t \sim 1$~\cite{wellein,braden}. Furthermore, the strength of the spin-phonon interaction for CuGeO$_3$
has been estimated to be $g \sim 0.1t$~\cite{werner}. 
Thus, the assumption of $t=0$ is not reasonable for CuGeO$_3$.
Although the spin-spin interaction is an important factor for the research on CuGeO$_3$~\cite{trebst}, 
the result in the present work shows that the phase transition occurs also in the system without the spin-spin interaction. 
The existence of the spin-spin interaction is not a necessary condition for the occurrence of the phase transition.

The magnetic properties in the spin system coupled to phonons could be investigated from knowledge of the frustrated systems.
Looking from a reverse viewpoint, 
we would be able to study the frustrated spin systems from knowledge of the spin systems coupled to phonons 
by quantum Monte Carlo simulation, the density matrix renormalization group method, and so on.

\begin{acknowledgment}

One of the authors (CY) acknowledges fruitful discussions with S. Todo. One part of the numerical calculations has been
performed on SGI Altix at ISSP, University of Tokyo.

\end{acknowledgment}

\end{document}